\RequirePackage{fix-cm}
\documentclass[twocolumn]{svjour3}          
\usepackage{graphicx}
\usepackage{amssymb,latexsym}
\usepackage{fancyhdr,amssymb}
\usepackage{url}
\begin{document}

\title{Parameter estimation by fixed point of function of information processing intensity}

\titlerunning{Parameter estimation by fixed point...}        

\author{Robert Jankowski         \and
        Marcin Makowski 		\and
        Edward W. Piotrowski
}


\institute{ R. Jankowski         \and
        M. Makowski 		\and
        E. W. Piotrowski \at
              Institute of Mathematics\\
              University of Bia\l{}ystok\\
              ul. Akademicka 2\\
              15-267 Bia\l{}ystok, Poland\\
							\email{rjankowski@math.uwb.edu.pl} (R. Jankowski)\\
              \email{makowski.m@gmail.com} (M. Makowski)\\
							\email{qmgames@gmail.com} (E.W. Piotrowski)
}

\date{Received: date / Accepted: date}

\maketitle

\begin{abstract}
We present a new method of estimating the dispersion of a distribution which is based on the surprising property of a function that measures information processing intensity. It turns out that this function has a maximum at its fixed point. We use a fixed-point equation to estimate the parameter of the distribution that is of interest to us. We illustrate the estimation method by using the example of an exponential distribution. The codes of programs that calculate the experimental values of the information processing intensity are presented. 
\keywords{Parameter estimation \and Fixed point theorem \and Information processing intensity}
\end{abstract}

\section{Introduction}
\label{intro}
In the case of effective collection of statistics, one of the priorities is to obtain information according to the principle of ``as much as possible as quickly as possible''. We analyse the process of capturing packets of any data in this context. We assume that the information measure of these chunks is a real random variable having a distribution with a finite dispersion. We are interested in characterising this type of distribution and estimating its dispersion. The method of capturing data packets (and also of making inferences about the distribution of information density for all of the packets based on measurement of these packets) should be optimal, i.e. it should only take into account the necessary part of such packets, i.e. those whose information measure is one-sidedly bounded by a certain constant a. The above assumptions lead to the development of a \textsl{function of information processing intensity} which has a surprising property, i.e. it has a maximum at its fixed point. We use a fixed-point equation to estimate the dispersion of a distribution.

Fixed-point theorems constitute a fascinating object of study. Since its beginnings, the development of fixed-point theory has been connected with its numerous applications in other fields of mathematics as well as in game theory and economics \cite{Fg,fa}. Naturally, fixed-point theorems are also used to learn about the distribution of a characteristic within a population based on data obtained from a sample, which is an incredibly important problem in the natural and social sciences \cite{Fg,Sb}. It is worth emphasising that the estimation method which is presented below can be interpreted in an intuitive, clear and interesting way, by means of maximising the function of information processing intensity, which corresponds to the effectiveness of this process.

\section{Information processing intensity}
\label{sec:1}
In order to illustrate this problem, let us use the example of an abstract relay $\mathfrak{A}$ which captures chunks of data (information). Information chunks that have an information value which is too low (or, analogously, too high) are rejected, whereas others are forwarded. This mechanism is analogous to the way in which a high-pass (or low-pass) filter functions. 
Let us assume that the information measure of information chunks that are captured by $\mathfrak{A}$ is a real random variable having a distribution that is characterised by parameter s. Let us also assume that the actions of capturing and rejecting a chunk (packet) of data as well as the process of forwarding the chunk of data that has been captured take the same amount of time.

We denote by $I_{in}$ a random variable which describe the value of information of each single non-filtered package. The probability density function of such a random variable is $pdf_{\gamma}(x)$ ($\gamma$ is the parameter of the distribution). $I_{out}$ is another random variable which measures the contents of the information package of intercepted (filtered) and transferred (emitted) information. We also need to define a random variable (takes values from the set of the natural numbers) $T$ which measures the total time of interception and transfer of the intercepted information package:
\begin{equation}
T=T_{in}+T_{out},\nonumber
\end{equation}
where $T_{in}$ is a random variable which represents the time of interception of one information package and $T_{out}$ represents the time of transfer of the same package. We assume that $E(T_{out})=1$, which means that the packages are emitted immediately.

Because of the above assumptions, \textbf{the measure of information processing intensity} is:

\begin{equation}\label{nat}
	\rho=\frac{\vert E(I_{out}-I_{in})\vert}{E(T)}.
\end{equation}
 We  calibrate the filter in a such way: $E(I_{in})=0$ (see Theorem \ref{thm} in next paragraph). We denote by $a$ the data cut-off (rejection) parameter (if the information measure of the information package is less than $a$ then such a package is rejected) \footnote{The number $a$ does not reflect any physical property. Our model is an abstraction of the investigated problem. An analogy can be discerned here to classical acoustic filters (electronic or mechanical). There exists a boundary frequency (amplitude) such that the fiter blocks the signal below (or above) it.}. 
We denote by $q$ the probability of rejection and by $p$ the probability of interception of the information package (of course, there is always $(p+q=1)$):\footnote{We assume that the symbol $[sentence]$ is equal to 1 if the sentence is true. In other cases it is equal to 0 (Iverson bracket).}
\begin{equation}\label{pr}
  \begin{array}{l}
	q:=E([x<a])=\int_{-\infty}^{a}pdf_{\gamma}(x+m) dx,\\\\
	p:=E([x\geq a])=\int_{a}^{\infty}pdf_{\gamma}(x+m) dx,
	\end{array}
\end{equation}
where $m$ is the first moment of random variable $I_{in}$. In the presented model, it is convenient to shift the domain of function $pdf_{\gamma}(x)$ so that the first moment of shift variable distribution is equal to 0.  We can calculate the expected time of interception of the information package:
$$
	E(T_{in})=(1-q) +2q(1-q)+3q^2(1-q) +\cdots=\frac{1}{1-q}.
$$
If we take into account the (\ref{pr}) and $E(T_{out})=1$ then it is easy to obtain:
\begin{equation}\label{czas}
	E(T)=1+E([x\geq a])^{-1}.
\end{equation}
The random variable  $I_{out}=[I_{in}\geq a]\,I_{in}$ has a probability density function $pdf_{\gamma}(x+m)$ which is cut off to the field $[a,\infty)$:
\begin{equation}\label{ppp}
	\frac{[x\geq a]}{E([x\geq a])}\,pdf_{\gamma}(x+m)\,.
\end{equation}
If we put an expected value of random variable $I_{out}$ and (\ref{czas}) to (\ref{nat}) then we get the formula for information processing intensity:

\begin{equation}\label{nate}
	\rho(pdf_{\gamma},a)_{right}:=\rho(a)=\frac{\int_{a}^{\infty} x\cdot pdf_{\gamma}(x+m)dx}{1+\int_{a}^{\infty}pdf_{\gamma}(x+m)dx}. 
\end{equation}

\section{Estimating a parameter of probability distribution}
\label{sec:2}
If we want to find the maximal value of functional (\ref{nate}) (with a chosen probability density function), we need to use the fixed-point theorem (c.f \cite{fix}):
\begin{theorem}\label{thm}
The maximal value $a^{*}$ of the function (\ref{nate}) lies at a fixed point of this function ($\rho(a^{*})=a^{*}$). Such point exist, is uniquely determined and $a^{*}>0$.
\noindent
\end{theorem}
The equation which allows to find the fixed point:
\begin{equation}\label{rownanie}
	\frac{\int_{a}^{\infty} x\cdot pdf_{\gamma}(x+m)dx}{1+\int_{a}^{\infty}pdf_{\gamma}(x+m)dx}=a,
\end{equation}
is used to determine the dependence of the parameter $\gamma$ of distribution $pdf_{\gamma}(x)$ on a fixed-point value $a^{*}$.

Given the above assumptions, let us note that the most effective estimation of the distribution parameter constitutes a greedy problem, i.e. the intensity of information processing is greater when the cut-off parameter (data rejection parameter) $a$ is moved closer to fixed point $a^{*}$ of intensity $\rho(a)$ in iterative process $\rho(a)\rightarrow a$. 
The $\rho$ function is a contraction in the neighbourhood of its fixed point. Where the modulus of the derivative of the probability density function is sufficiently high (when $|\rho'|>1$)  intensity $\rho$ is not a contraction. However, for every typical probability density function\footnote{With finite value of the first two moments, and eg placed on the list \url{ http://en.wikipedia.org/wiki/List_of_probability_distributions.}} we can reach a fixed point through iteration:
\begin{equation}\label{iteracja}
a_{k+1}=\rho(a_k),
\end{equation}
by starting with any $a>0$ (Banach contraction mapping principle). 

Now, if we put such empirical value $a_{eff}^{*}$ into the fixed-point equation (\ref{rownanie}), we can calculate an estimated value of parameter $\gamma_{eff}$ from distribution $pdf_{\gamma}(x)$. Theorem \ref{thm} is true for any continuous probability distribution. That is why the method of estimating parameters (which is described above) is effective for such distributions. 

If we analyse the analogical model (in which packages with a too large value of information are rejected), we obtain symmetric conclusions. The equation which describe information processing intensity in such a model takes the form:

\begin{equation}\label{nateleft}
	\rho(pdf_{\gamma},b)_{\normalfont left}:=\rho(b)=\frac{-\int_{-\infty}^{-b} x\cdot pdf_{\gamma}(x+m)dx}{1+\int_{-\infty}^{-b}pdf_{\gamma}(x+m)dx}. 
\end{equation}
Also, in such a model the maximal value of function (\ref{nateleft}) lies at a fixed point of this function. Like before equation of the fixed point of function (\ref{nateleft}) can be used to estimate parameters of distributions. From now we denote by the \textsl{right-fixed point} the fixed point of function (\ref{nate}) and by the \textsl{left-fixed point} the fixed point of function (\ref{nateleft}). 
In the next section we present an example of using such fixed points to estimate the parameter of exponential distribution. 

\section{Example of fixed point method}
\label{sec:3}
Let us consider the probability density function of exponential distribution:
\begin{equation}\label{wykn}
pdf_{\gamma}(x)=\gamma\, e^{-\gamma\, x},
\end{equation}
$x\in [0,\infty)$ and $\gamma >0$. If  $m=\gamma^{-1}$ is the expected value of the random variable of distribution (\ref{wykn}), then: 

\begin{equation}\label{wykladniczy}
\widetilde{pdf_{\gamma}}(x):=pdf_{\gamma}(x+m)=\gamma\, e^{-\gamma\, x-1}
	\end{equation}
	where $\gamma > 0$, $x> -1/\gamma$. The first moment for such probability distribution is equal to zero.
	
Let us see the results of applying the procedure as described above (for the right and left-fixed point) in the computer experiment. For that purpose a sequence of 1000 values of exponential distribution (\ref{wykn}) with $\gamma~=~1$ was generated.	

\subsection{Estimation using the right fixed point} 
\label{sec:4}
After integrating (\ref{rownanie}) with probability density function (\ref{wykladniczy}), we obtain the following fixed-point equation:
\begin{equation}
\frac{a\,\gamma +1}{\gamma \,{e}^{a\,\gamma +1}+\gamma}=a\,.\nonumber
\end{equation}
We can rewrite the above equation as:
\begin{equation}\label{defaqa}
	a=\frac{{e}^{-a\,\gamma -1}}{\gamma}.\nonumber
\end{equation}

The empirical value of fixed point $a_{eff}^{*}$ allows to find an estimated value of the parameter:
\begin{equation}\label{estw}
\gamma_{eff}=\frac{W(e^{-1})}{a_{eff}^{*}},
\end{equation}
where $W(z)$ is the Lambert W function. The estimated dispersion of distribution (which is a measure by variance) is equal to $\gamma_{eff}^{-2}$.

If we use the exponential distribution with parameter $\gamma=1$, then the right fixed point (the maximum) of function (\ref{nate}) is equal to 0.278465. The right experimental (see appendix)  fixed point (found after the 5th iteration) is equal to:
\begin{displaymath}
a_{eff}^*=0.281203\,.
\end{displaymath} 
Therefore, the estimated value of the parameter (according to formula (\ref{estw})) is equal to:
\begin{displaymath}
\gamma_{eff}=0.990261\,.
\end{displaymath}

The relative error of the estimated value of the parameter (with all of the above assumptions) is equal to about 0.97\%.
\begin{figure}[!ht]
  \centering
    \includegraphics[width=3in]{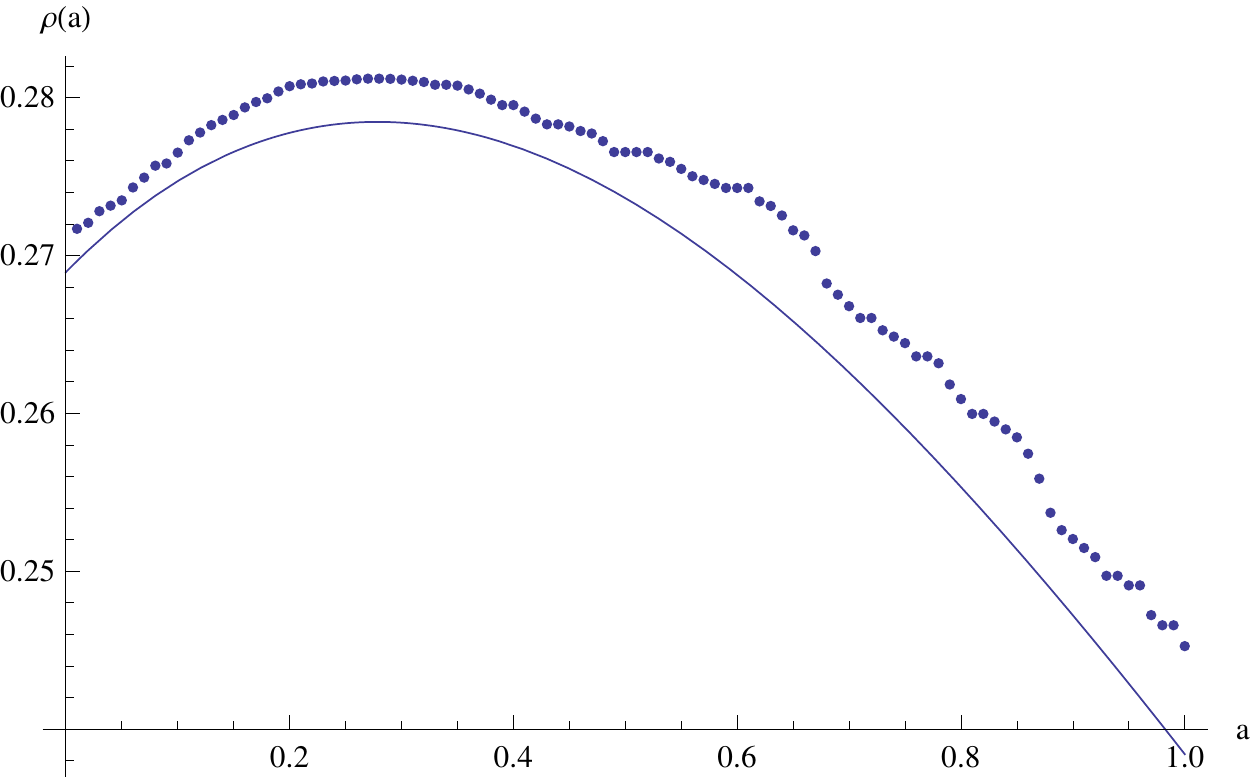}
    \caption{The part of the graph of theoretical (continuous line) and experimental (points) right intensity function $\rho(a)$,  $a\in[0,1]$.}
    \label{rhoR}
\end{figure}
Figure \ref{rhoR} shows a theoretical and experimental graph of the right intensity function $\rho(a)$ from the discussed example.

\subsection{Estimation using the left fixed point}
\label{sec:5}

\begin{figure}[!ht]
  \centering
    \includegraphics[width=3in]{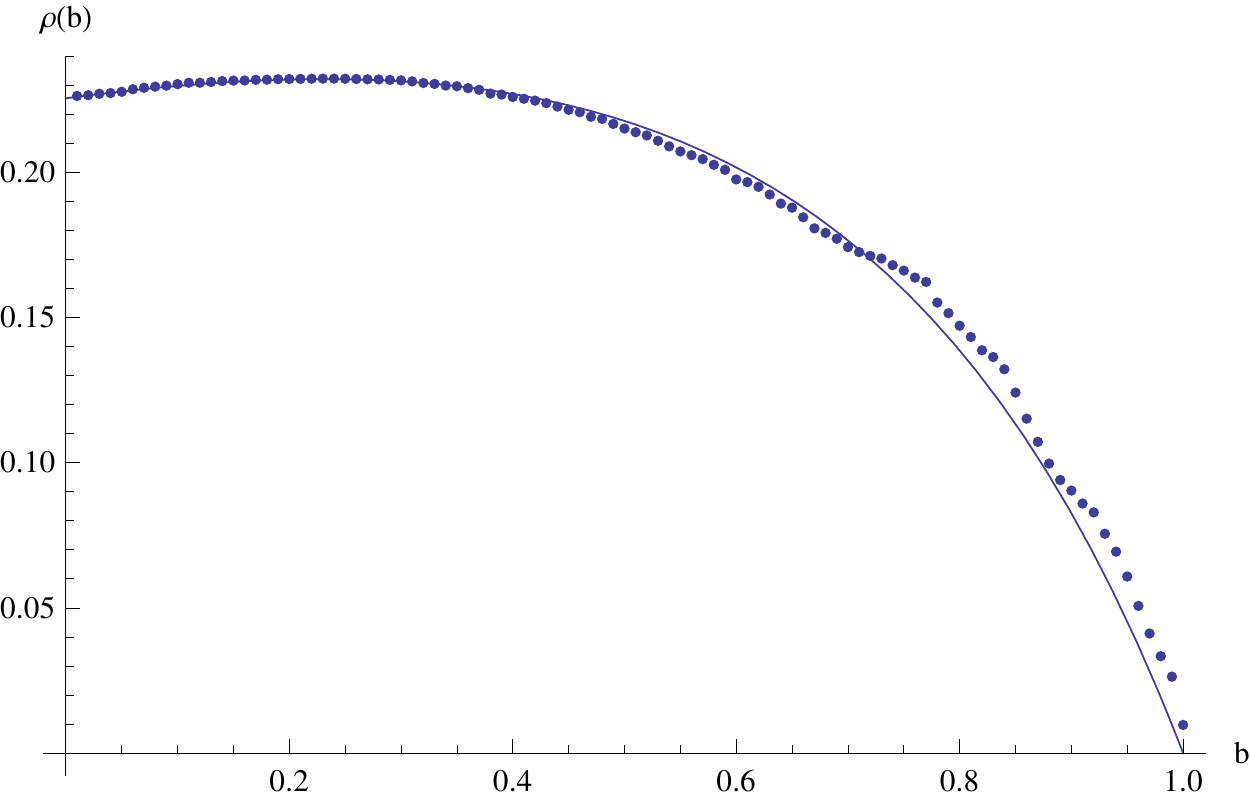}
    \caption{The part of the graph of theoretical (continuous line) and experimental (points) left intensity function $\rho(b)$,  $b\in[0,1]$.}
    \label{rhoL}
\end{figure}

Let us see the result of estimating the parameter (with all of the experiment’s  assumptions) by means of the left-fixed point. We denote by $b_{eff}^*$ the experimental left fixed point. Analogical calculations (just as for the right-fixed point) shows that: 
 \begin{equation}\label{estlew}
\gamma_{eff}=-\frac{W((-2e)^{-1})}{b_{eff}^{*}}.
\end{equation}
The theoretical value of such a left fixed point is equal to \newline
0.231961. The experimental value (also after the 5th iteration) is equal:
\begin{displaymath}
b_{eff}^{*}=0.232172\,,
\end{displaymath}
and (see (\ref{estlew})):
\begin{displaymath}
\gamma_{eff}=0.999089\,.
\end{displaymath}
The experimental and theoretical graph of left function $\rho(b)$ is presented in Figure \ref{rhoL}. 
Estimation by means of the left fixed point gives the relative error at a level of about 0.09\%\,.

\section{Conclusion}
\label{sec:6}
In this paper we analysed a new method of estimating a distribution parameter by using the function of information processing intensity as well as its property, i.e. a maximum at the fixed point. Theorem \ref{thm} also has other interesting applications \cite{g,rQ,rT,sy,fix2}.

What is characteristic of the method we have presented is that when calculating the function of information processing intensity (both on the left and on the right), only those values which meet the condition specified by the cut-off parameter are used. Therefore, only a part of the available data influences the final result. It is in this context that the result which was obtained for the example described here, i.e. one related to estimating the parameter of an exponential distribution, seems to be interesting. The arithmetic mean of the generated sample of 1000 values with respect to this distribution with parameter $\gamma = 1$ equals 1.010836, which means that the relative error of this value is approx.  1.08\%. We estimated the parameter in question with greater accuracy both for the right and left fixed point by using only a part of the data. 

The codes of programs that calculate the experimental values of the $\rho$ function are presented in the appendices.
\begin{acknowledgements}
This work  was supported by the Polish National Science Centre
under the project number \textbf{DEC-2011/01/B/ST6/07197}.
\end{acknowledgements}

\appendix
\section{Program code which calculate experimental value of information processing intensity}
\label{sec:7}
Let us show (in Mathematica programming language) program codes which allows to find experimental value of information processing intensity. In both codes $z$ is a list of values of the random variable $I_{in}$.

\subsection{Right $\rho$} 
 \begin{verbatim}
 n:10 000;
m = Apply[Plus, z]/n;
rho[a_]:= 
  Module[{o = 0, numerator = 0., denominator = 0}, 
   Do[If[z[[j]] > m + a, 
	     numerator = numerator + z[[j]] - m; 
     denominator = denominator + j - o + 1;
     o = j], {j, 1, n}];
   If[ denominator == 0, denominator = 1];
   numerator/denominator];
  \end{verbatim}
  \subsection{Left $\rho$}
  \begin{verbatim}
 n:10 000;
m = Apply[Plus, z]/n;
rho[b_]:=
  Module[{o = 0, numerator = 0., denominator = 0}, 
   Do[If[z[[j]] < m - b, 
	     numerator = numerator + z[[j]] - m; 
     denominator = denominator + j - o + 1;
     o = j], {j, 1, n}];
   If[ denominator == 0, denominator = 1];
   -numerator/denominator];
  \end{verbatim}



\end{document}